\documentclass[10pt]{article}

\usepackage{graphicx}

\title{On the Aging Dynamics in an Immune Network Model}

\author{Mauro Copelli, Rita Maria Zorzenon dos Santos  \\
\small Laborat\'orio de F\'{\i}sica Te\'orica e Computacional \\ 
\small Departamento de F\'\i sica, Universidade Federal de Pernambuco \\
\small Cidade Universit\'aria, 50670-901, Recife, PE, Brazil \\
\ \\
Daniel Adri\'an Stariolo\footnote{Research Associate of the Abdus
Salam International Center for Theoretical Physics, Strada Costiera
11, Trieste, Italy} \\
\small Instituto de F\'\i sica, Universidade Federal do Rio Grande do Sul\\
\small CP 15051, 91501-970, Porto Alegre, Brazil}

\date{\today}

\begin{document}

\maketitle

\abstract{Recently we have used a cellular automata model which
describes the dynamics of a multi-connected network to reproduce the
refractory behavior and aging effects obtained in immunization
experiments performed with mice when subjected to multiple
perturbations. In this paper we investigate the similarities between
the aging dynamics observed in this multi-connected network and the
one observed in glassy systems, by using the usual tools applied to
analyze the latter. An interesting feature we show here, is that the
model reproduces the biological aspects observed in the experiments
{\it during} the long transient time it takes to reach the stationary
state. Depending on the initial conditions, and without any
perturbation, the system may reach one of a family of long-period
attractors. The perturbations may drive the system from its natural
attractor to other attractors of the same family. We discuss the
different roles played by the small random perturbations (``noise'')
and by the large periodic perturbations (``immunizations'').}

\section{Introduction}
\label{sec:intro}

In this paper we discuss a model for the evolution of the immune
repertoire of B cells, which are responsible for the humoral immune
responses.  B cells belong to one of the main classes of white blood
cells: the lymphocytes. These cells carry on their surface the order
of $10^5$ molecular receptors (proteins) and once activated they
produce antibodies, which are copies of their molecular
receptor. During the life time of an individual the immune system is
able to produce the order of $10^{11}$ different antibodies or
different populations of B cells. The antigen (virus, bacteria,
poison, cellular residue, etc) is not recognized as a whole but by its
epitopes, which are patches on its structure that may be recognized by
specific sites of the antibody molecules. By pattern recognition
different antibodies will mark the epitopes of a given antigen,
therefore forming a complex that will be eliminated by macrophages
(another class of white blood cells)~\cite{janeway,nelson}.

According to clonal selection theory~\cite{nelson,janeway}, elements
that challenge the immune system will determine the populations
(clones) of B cells that will proliferate: those populations will
produce antibodies which will be able to recognize different epitopes
of the specific antigen. The immune network
theory~\cite{Jerne,nelson}, however, is based on the fact that the
antibodies (and molecular receptors) are able to recognize and to be
recognized, and therefore during the immune response there are
different types of interaction: antigen-antibodies and antibodies-B
cells. In other words, when a given population of B cells is activated
by the presence of a given antigen the produced antibodies will not
only mark the specific antigens but also activate new B cell
populations with complementary molecular receptors, which in turn will
recognize them. The increase on the concentration of these
complementary populations, on their turn, will maintain the
proliferation of the antigen recognizing population, installing a
feedback mechanism that will keep several populations activated.  This
kind of dynamics will generate a functional multi-connected network
among different populations of B cells that will be dynamically
regulated by mechanisms of activation and suppression. The network
will then play an important role on the regulation of the immune
responses. Although the immune network theory is part of the
current immunological thinking, there are only few experiments
supporting the interaction among clones with complementary receptors
and the existence of such a
network~\cite{holmberg,coutinho89}. According to these experimental
findings, if the network exists only $10-20\%$ of the populations will
belong to it, the rest of the immunocompetent populations remaining at
rest.

Recently we have successfully used a mathematical
model~\cite{sw,zsb95,bzs97,zs99} (inspired in a previous one proposed
by de Boer, Segel and Perelson~\cite{bsp}) which takes into account
the main features of Jerne's immune network theory, to simulate
experiments on immunization and aging performed with mice~\cite{zsb98}
that could not be explained by the clonal selection theory . The
simulations allowed to interpret the experimental results from the
point of view of the immune network theory.

The model allows to follow the evolution of the concentrations of the
different populations of B cells in discrete shape space, a formalism
which maps all possible molecular receptors of a given organism into
points of a $d$-dimensional space. To each point (receptor) we
associate a clone that corresponds to the population of B cells and
antibodies characterized by this molecular receptor. The concentration
of each clone will be described by a three-state automaton
representing low, medium and high concentrations, and the interaction
among different clones is based on complementarity. As far as we know,
the model~\cite{bsp,sw,zs93,zsb95,bzs97} corresponds to the first
successful attempt in describing the dynamics of the immune network as
proposed by Jerne~\cite{Jerne} and recently it could be used to
reproduce experimental results performed with mice~\cite{zsb98}.
Besides the biological implications of the results obtained in
Ref.~\cite{zsb98}, the dynamical behavior exhibited by this model in
the biologically relevant parameter region is quite interesting by
itself and should be better investigated. This region has been
shown~\cite{zsb95} to exist for dimensions $d\geq 2$ and comprises a
broad stripe near the transition between stable and chaotic behaviors,
in which the model describes a multi-connected functional
network~\cite{zsb95,bzs97}. In this region we found the majority of
the populations in the resting state (low concentration) while the
activated ones may reach $10-20\%$ of the total number of
populations. The activated populations are aggregated in clusters of
different sizes, which fuse and split as time passes following an
aggregation and disaggregation dynamics. Therefore the largest cluster
at each time step is found in a different region~\cite{bzs97}.

In the experiments described in ref.~\cite{zsb98}, 6 mice of the same
linage are subjected to the immunization protocol as follows: the
researchers inject ova by means of an intra-venal injection, wait for
14 days and measure the number of specific antibodies. Then they
inject again the same amount of antigen, wait for 7 days and measure
the amount of antibodies, and continue by repeating the same protocol
every 7 days.  In order to simulate the immunization protocol, the CA
is subjected to specific perturbations by flipping chosen resting
populations (low concentration) to their activated state (high
concentration).  Depending on the perturbation (damage) size, it may
disappear after a few time steps or part of the damage (activated
populations) may be incorporated to the network. In other words, the
memory about the perturbations is due to the ability of the system to
adapt (plasticity) and incorporate information about them. We have
used two types of perturbations: random small ones, which correspond
to the noise that mice are subjected to during the experiment, caused
by the environment, and periodic large ones that will simulate the
immunization protocol of multiple antigen perturbations under the same
conditions~\cite{zsb98}. After few presentations there is a saturation
of the response of the system to the perturbation.  During this
process, the system incorporates new information and some of the
previous information is lost, keeping the number of activated sites in
the network almost unchanged. This kind of behavior was also observed
experimentally in mice~\cite{bruno97}, where saturation is related to
a refractory behavior of the immune system.

We have also observed aging effects on the dynamics of this
system~\cite{zsb98}. An older system is more rigid: the network loses
plasticity to incorporate new information~\cite{zsb98,ana97,lahmann}.
A recent study~\cite{bzs2001} has shown that the distribution of
cluster sizes during the time evolution of the system has a
characteristic cluster size (exponential behavior), but the
distribution of persistence times (the period during which a given
population remains activated and belongs to the network) exhibits a
power law behavior.  While the existence of a characteristic cluster
size may be related to the loss of plasticity, the power law behavior
of the persistence time may be associated to the memory generated by
the dynamics of the system.

The question we address here refers to how the learning process takes
place dynamically and what is the cause of the loss of plasticity. The
slow dynamics observed in this system presents analogies with the
physical aging effects observed and reported on glass
studies~\cite{struik,bocukume}: as the system gets older (ages) there
is a loss of plasticity for structural or molecular relaxation and
less changes are observed during the relaxation time.  The mechanisms
underlying the slow dynamics of glassy systems are the spatial (or
geometric) disorder related with the difficulty to satisfy
simultaneously all microscopic interactions, a characteristic called
{\em frustration}. Glasses and spin glasses exhibit a rough energy
landscape with many local minima which is responsible for slowing down
the relaxation towards equilibrium.  Their dynamics depends on the
past history of the sample and under small perturbations they relax
slowly. In our biologically motivated model the non-local interactions
are based on complementarity (both perfect and slightly defective
matches) and reflect the activation and suppression mechanisms in the
complementary regions of the shape space. The analogue to frustration,
in the network, is generated by the inability of the system to
incorporate information and to satisfy the constraints of the
activation and suppression mechanisms.  We show that the CA dynamics
gives rise to a large family of periodic attractors which are very
robust, and could be regarded as the analogues of the minima in the
glassy systems. The lost of plasticity and aging effects will
therefore be related to the non-ergodicity in the phase space.

In this paper the dynamics of the immunological responses in the
network model are investigated using the common tools adopted in the
study of glassy systems~\cite{bocukume}. In particular we will focus
on the relaxation of two-time autocorrelations of the B-cell
populations, the structure of the attractors and effects of
perturbations and immunizations. The paper is organized as follows: in
section~\ref{sec:themodel} we describe the model and the procedures
adopted in order to measure the correlations among different
configurations of the system; in section~\ref{sec:results} we present
and discuss the results obtained for small and large perturbations and
in section~\ref{sec:discussion} we present a summary, prospectives for
future work and some conclusions.

\section{The Model}
\label{sec:themodel}

The model under study here is a modified
version~\cite{zsb95,bzs97,zs99} of the model proposed by Stauffer and
Weisbuch~\cite{sw}, which in turn was inspired in a previous model
proposed by de Boer, Segel and Perelson~\cite{bsp} to describe the
time evolution of the immune repertoire. It is a deterministic window
cellular automaton model based on the shape space
formalism~\cite{peros}, which describes the interactions of
B-lymphocytes and antibodies, and the main mechanism underlying these
interactions, which is pattern recognition (lock-key interaction). The
dynamics of the system is regulated by specific interactions involving
complementary molecular receptors of the different clones. The memory
about the relevant antigens, presented to the system during its past
history, emerges from the dynamics of the system, rather than being
stored in a static registry.

To each point of a $d$-dimensional discrete lattice we associate a
given receptor, which in turn will be described by $d$-coordinates
representing important physical-chemical features of the receptor that
differentiate one from the other~\cite{peros}. Since clones are
classified according to their molecular receptor, to each point $\vec
r$ of the discrete shape space we associate a three-state automaton
$B(\vec r,t)$ that will describe the concentration of its population
over the time: low ($B(\vec r,t)=0$), intermediate ($B(\vec r,t)=1$)
and high ($B(\vec r,t)=2$).

The time evolution of the cellular automaton is based in a non-local
rule: population $B(\vec r,t)$ at site $\vec r$ is influenced by the
populations at site $-\vec r$ (its mirror image or complementary
shape) and its nearest-neighbors ($-\vec r + \delta \vec r$)
(representing defective lock-key interactions). The influence on the
population at site $\vec r$ due to its complementary populations is
described by the field $h(\vec r,t)$:

\begin{equation}
h(\vec r,t)=\sum_{ \vec r^{\,\prime}\in (-\vec r + \delta \vec r)}
B( \vec{r}^{\, \prime},t)
\end{equation}
where for a given $\vec r$ the sum runs over the complementary shape $
\vec r^{\, \prime} = -{\vec r}$ and its nearest neighbors. Due to the
finite number of states of the population $B$, the maximum value of
the field $h( \vec r)$ is $h_{max}=2(2d+1)$. The updating rule is
based on a window of activation, which describes the dose response
function involved in B-cell activation~\cite{zsb95,sw,bsp}. There is a
minimum field necessary to activate the proliferation of the receptor
populations ($\theta_1$), but for high doses of activation (greater
than $\theta_2$) the proliferation is suppressed. The updating rule
may be summarized as:

\begin{equation} 
\label{eq:dyn}
B( \vec r,t+1)= \left\{ \begin{array}{ll}
B( \vec r,t) +1 \qquad & {\rm if} \qquad \theta_1 \leq h( \vec r,t) \leq \theta_2 \\
B(\vec r,t) - 1 &{\rm otherwise} \end{array} \right.
\end{equation}
but no change is made if it would lead to $B=-1$ or $B=3$.  We define
the densities of sites in state $i$ at time $t$ as
$B_i(t)$ ($i=1$, 2, 3).

The initial configurations are randomly generated according to the
following concentrations: $B_1(0)= B_2(0)=x/2$, while the remaining
$L^d(1-x)$ sites are initiated with $B(\vec r,0)=0$. This model may exhibit
stable or chaotic behaviors depending on the values of $x$, $\theta_1$
and $\theta_2-\theta_1$.  However it is on the transition region
between the two behaviors that the model behaves like a
multi-connected network~\cite{zsb95}.

In order to simulate the immunization protocol performed in the mice
experiments we have followed the procedure which is described in
Ref.~\cite{zsb98} and summarized below. We have adopted the scale of~5
time steps corresponding to~1 day~\cite{zsb98}.  While the system
evolves according to the deterministic dynamics (eq.~\ref{eq:dyn}),
small and large perturbations can be produced, by setting the state of
the chosen sites at $B(\vec r, t)=2$.

\paragraph{Small Perturbations} 
The small perturbations account for the immunological stimuli (noise)
coming from the environment.  The time interval between two
consecutive small perturbations is a random number uniformly
distributed between 1 and 100 time steps. Each perturbation
corresponds to a random number of damages (from 1 to 3) introduced at
regions of resting populations ($B=0$) which are randomly drawn (at
every perturbation). The size of each damage may also vary randomly
from 1 to 3 (onion-like) concentric layers around a central site
(containing 7, 25 or 63 populations, respectively, in 3 dimensions).

\paragraph{Large Perturbations}
The large perturbations correspond to the immunization protocol which
starts at a predetermined age of the mice. Like in the experiments, we
stimulate the system periodically (every 35 steps $\simeq$ 1 week),
and always in the same region (which is initially chosen at random but
kept unchanged along the simulation). The damage size in this case
corresponds to six layers (377 populations) around an specific site.

Previous results on this model have shown~\cite{zsb98} that the
response to the immunizations presents a strong dependence on the
initial time (``age'') at which the periodic protocol starts (fitting
experimental data extremely well for mice whose immunization protocol
started at different ages). This has led us to the conjecture that the
dynamical behavior of the system might be at least qualitatively
similar to that of some glassy systems, despite the non-Hamiltonian
nature of the CA dynamics.

One of the quantities commonly used in the study of glassy systems is
the two-time autocorrelation function between the system
configurations at two given times $t$ and $t_w$. A common experiment
in glassy systems consists in preparing the system at a high
temperature and suddenly making a quench to a low temperature. Then
the system is allowed to relax up to a {\em waiting time $t_w$}, whose
configuration is recorded. As the system continues to relax the
autocorrelations between the instantaneous configurations at time $t >
t_w$ and that at time $t_w$ are computed. The waiting time $t_w$ is
called the {\em age} of the system in context of glassy systems. The
monitoring of the two-time autocorrelations gives important insights
on the relaxation process. The dynamics, whether stationary or not,
can be readily recognized on the $t_w$ dependence of the
autocorrelations, since in a stationary process two-time quantities
depend only on time differences. Consequently by plotting correlations
as a function of time difference $t-t_w$ for different waiting times
it is possible to distinguish between an essential out of equilibrium
process from a stationary one.  The aging processes observed in glassy
systems are then related to the lack of temporal
invariance~\cite{bocukume}.

Inspired on this approach, we will analyze the multi-connected network
dynamics by defining and analyzing quantities analogous to the
two-time autocorrelations for the CA:

\begin{eqnarray}
C_{tot}(t,t_w) & = & \frac{1}{N} \sum_{\vec r}^N \delta\left( B(\vec
r,t_{w}), B(\vec r,t)\right) \\ C_{22}(t,t_w) & = &
\frac{1}{B_2(t_{w})N} \sum_{\vec r | B(\vec r,t_w)=2}^N \delta\left(
B(\vec r,t), 2\right),
\end{eqnarray}
where $N=L^d$ is the total number of populations in the
system. $C_{tot}$ and $C_{22}$ amount to normalized proximities (using
Hamming distance as a measure) and from now on will be referred to
simply as correlation functions. These quantities will be analyzed for
different protocols: without any perturbation, with small
perturbations (noise) and with large perturbations (immunization
protocol) in order to differentiate the effects of the different kinds
of perturbations.

A complementary view of the long-term behavior of the system can be
obtained by looking at the attractors to which the system evolves. In
the present case it will consist mainly of cycles, which will be
reflected by the periodicity of the correlations. For this purpose, we
have obtained the return maps of the consecutive maxima of the
correlation functions. Note that the period of the maxima thus
obtained does not correspond to the period of the real system, which
is at least twice as long~\cite{zs93}.

\section{Results}
\label{sec:results}

\subsection{Without perturbations}

In previous works~\cite{zs93,zsb95,bzs97} it has been shown that
depending on the initial concentration $x$ of active populations the
system may exhibit periodic or chaotic behavior. Systems with low
initial concentrations of active sites ($x<x_c$) evolve to either
fixed points or orbits with short periods, while for $x>x_c$ chaotic
attractors appear. However, the biologically relevant region is in the
transition region between these two behaviors, where the system
reaches one of several periodic orbits (as will be seen below) with a
very long period and after a long transient. From now on, all the
results have been obtained using the same parameters adopted in
Ref.~\cite{zsb98}: $d=3$, $\theta_1=h_{max}/3$, $\theta_2=2h_{max}/3$
and $x = 0.26$ (on the transition region).

\begin{figure}[!tb]
\begin{center}
\includegraphics[width = 12cm,height=8cm]{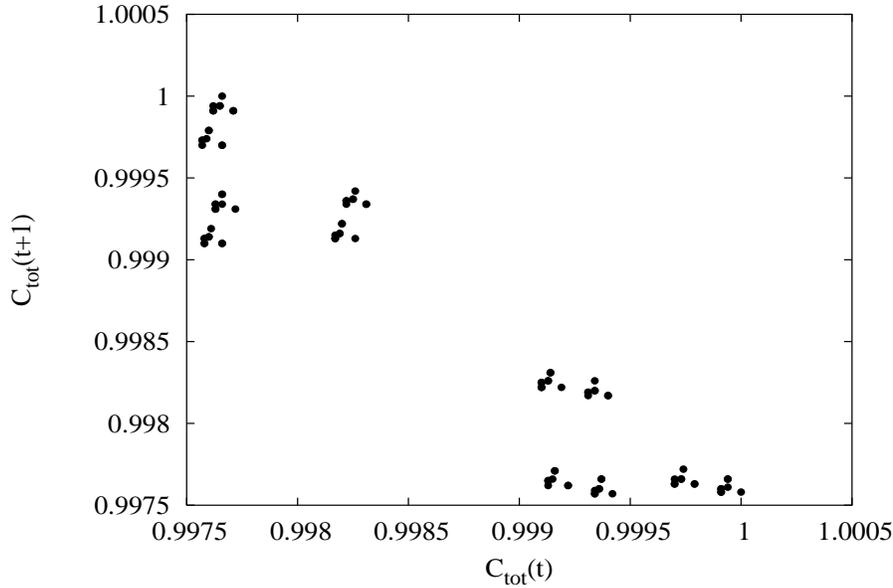}
\caption{Return map for maxima of the total correlation function $C_{tot}(t)$ for
$L=50$ and $t_w=10000$ (single run).}
\label{fig:x26r}
\end{center}
\end{figure}

Without any perturbation, the system evolves after a transient time
towards a cycle with a large period, as shown in the return map of
Fig.~\ref{fig:x26r}. We have also varied the waiting time from $10$ to
$100000$ (not shown). The greater the waiting time, the closer to the
attractor the system is, which is revealed by larger values of the
autocorrelation functions. Once $t_w$ is larger than the transient,
the time series for $C_{tot}(t,t_w)$ (and also $C_{22}(t,t_w)$ will
include unity (see Fig.~\ref{fig:x26r}) and will not change for larger
values of $t_{w}$.

\begin{figure}[!tb]
\begin{center}
\includegraphics[width =8cm,height=12cm, angle=-90]{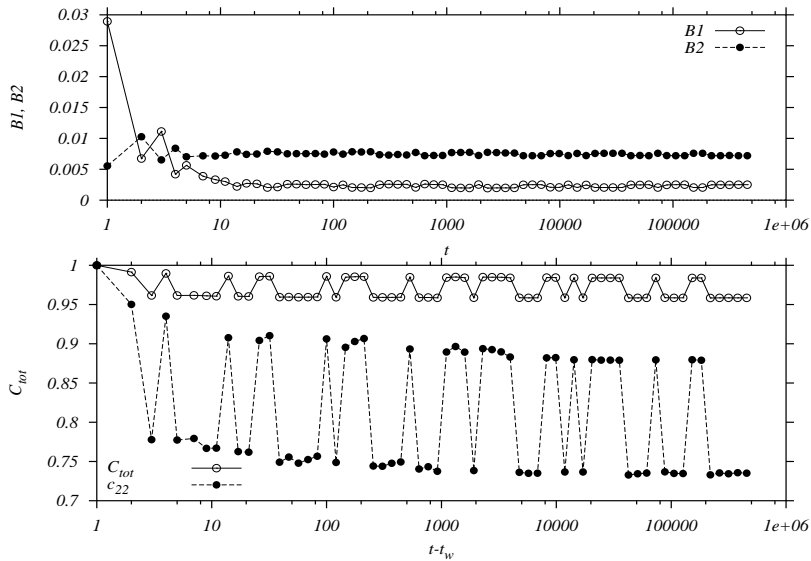}
\caption{Densities of intermediate and high concentrations vs. time (upper panel) and autocorrelations vs. $t-
t_w$ for $L=50$ and $t_w=100$, without any perturbation.}
\label{fig:L50np}
\end{center}
\end{figure}

A typical result for the time evolution of the correlation functions
for $t_{w}=100$ is shown in Fig.~\ref{fig:L50np}. The upper panel
corresponds to the time evolution of the densities $B_1$ (intermediate
concentrations) and $B_2$ (high concentrations), while the lower panel
corresponds to $C_{tot}(t,t_w)$ (open circles) and $C_{22}(t,t_w)$
(filled circles). Notice that while the concentrations relax to
approximately constant values in a short time, $C_{tot}$ and $C_{22}$
take much longer to reach their attractors (note the logarithmic time
scale). In order to study how the system relaxes towards the attractor
it is more convenient to make use of $C_{22}$, since it measures the
changes in the (more relevant) network of activated populations. In
the case shown in Fig.~\ref{fig:L50np} the transient time needed to
attain the attractor is ${\cal O}(10^4)$ steps. It is necessary to
point out, however, that this typical relaxation time is important
only for the physical aspects of the dynamics. When mapped into the
biological problem, it would correspond to $\sim 5.5$ years, which is
much longer than the average life time of the mice used in the kind of
experiment we simulated. Therefore the relevant behavior, from the
biological point of view, happens to be in the transient of the model
and not in its stationary state, a result which is interesting on its
own.

\subsection{Random Small Perturbations}
\label{sec:smallpert}

How does the behavior of the system change when random small
perturbations are produced on the parameter region used to simulate
the real experiments performed with mice~\cite{zsb98}? In order to
investigate this issue, only the small perturbations described in
Section~\ref{sec:themodel} were produced, starting at time zero.

\begin{figure}[!bt]
\begin{center}
\includegraphics[width = 0.5\textwidth, angle=-90]{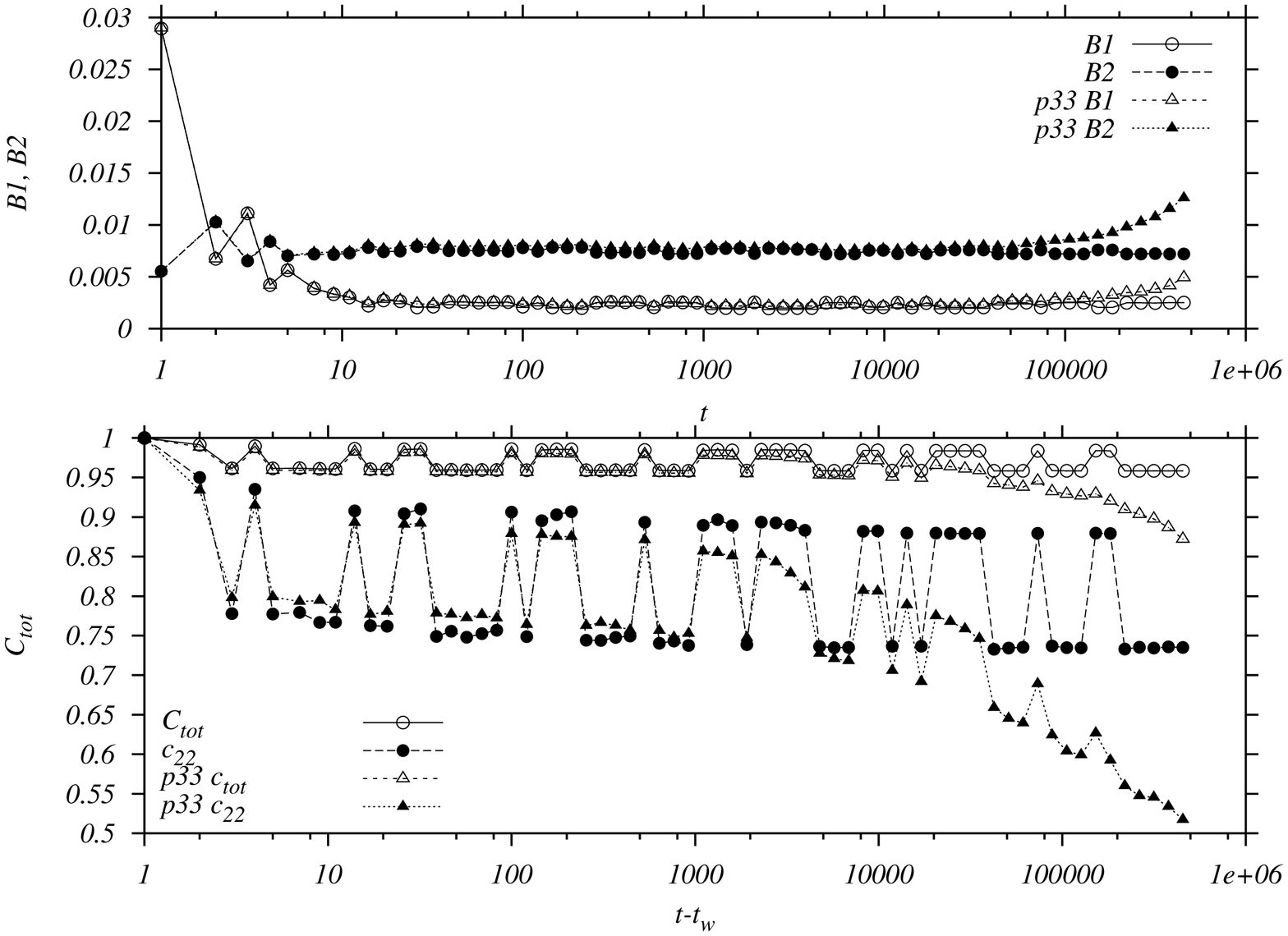}
\caption{Densities vs. time (upper panel) and correlations vs. $t-
t_w$ for $L=50$ and $t_w=100$; without perturbation (circles) and with
small perturbations (triangles).}
\label{fig:L50p33}
\end{center}
\end{figure}
\begin{figure}[!tb]
\begin{center}
\includegraphics[width = 8cm,height=12cm, angle=-90]{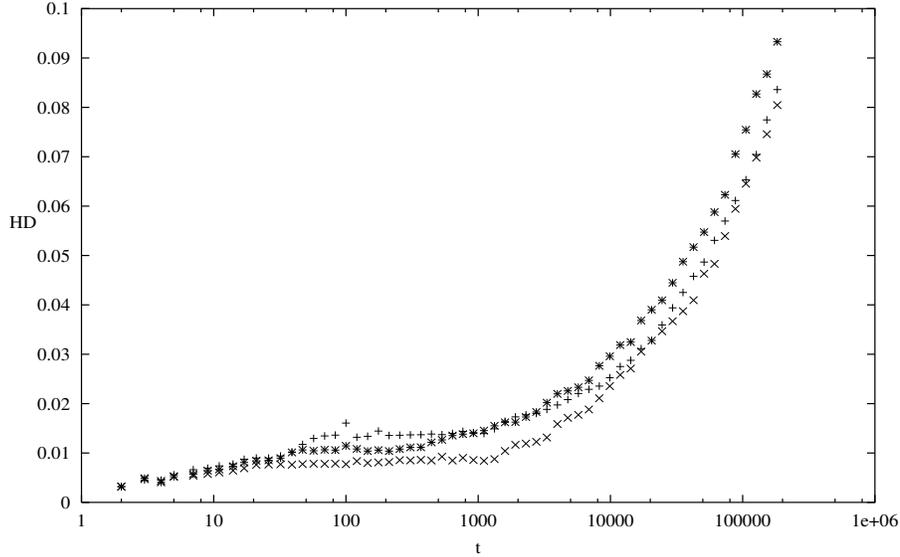}
\caption{Normalized Hamming distance between two initially identical
configurations subjected to different sequences of small perturbations
(three samples).}
\label{fig:dano}
\end{center}
\end{figure}

In Fig.~\ref{fig:L50p33} we compare the time evolution of the
densities and correlation functions obtained for single runs for the
system without perturbations and with small random perturbations. Note
that in the case with small perturbations the correlations initially
follow those of the purely deterministic system, showing only small
differences. After about $10^3-10^4$ steps, however, they start
decreasing faster, indicating some sort of cumulative effect that
drives the system away from the region in phase space that it had
approached until $t_w$. These effects are more easily noted for
$C_{22}$.  Moreover, the perturbations do not alter the behavior of
the densities, as expected, since the number of activated populations
is kept approximately constant by the self-regulatory mechanisms
embedded in the dynamical rules. The changes are observed only in the
populations that belong to the active network: in order to incorporate
new information (new populations), older ones are deactivated. From
the biological point of view, the realizations of the perturbations
(small and large) differ from one individual to another, building the
identity of each individual. At the end of life each individual will
have a different history in terms of perturbations (antigen
presentations) translated into the configuration of the populations
belonging to the active network.  This is nicely illustrated by
Fig.~\ref{fig:dano}, where two initially identical copies of a system
undergo different realizations of the small perturbations according to
the protocol described in section~\ref{sec:themodel}. The Hamming
distance between them grows on a long time scale, revealing the
mechanisms behind the behavior of the correlations in
Fig.~\ref{fig:L50p33}.

\begin{figure}[!tb]
\begin{center}
\includegraphics[width = 0.5\textwidth, angle=-90]{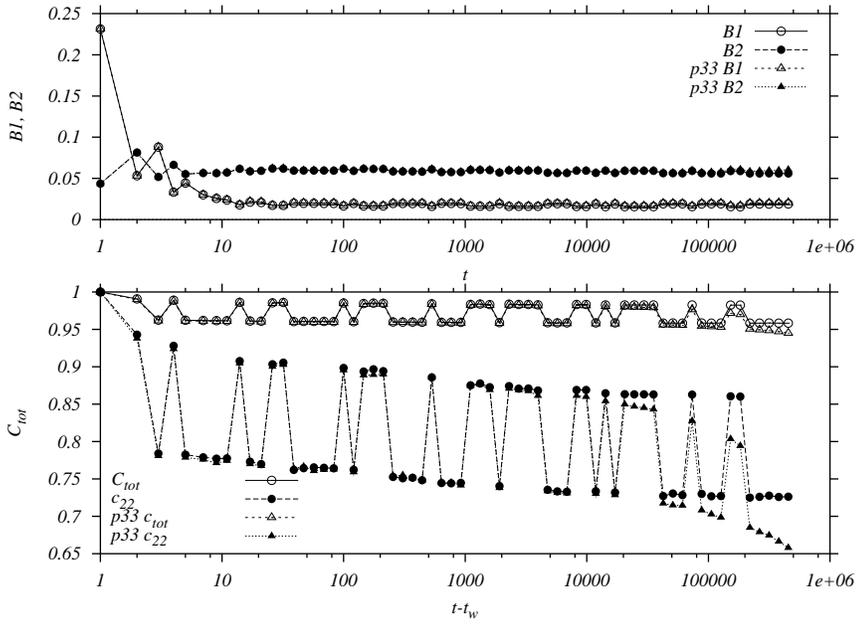}
\caption{Densities vs. time (upper panel) and autocorrelations vs. $t-
t_w$ for $L=100$ and $t_w=100$ (the same conditions used in
Fig.~\ref{fig:L50p33}); without perturbation (circles) and with small
perturbations (triangles).}
\label{fig:L100p33}
\end{center}
\end{figure}

Returning to Fig.~\ref{fig:L50p33}, it is important to stress that the
changes observed in the concentrations for very long times (upper
panel) are due to finite size effects. They are caused by the fact
that all small perturbations are produced on regions of resting
populations. For a finite system, after a long time all the
possibilities will have already been explored. Increasing the size of
the system, the changes on the densities disappear. This is shown in
Fig.~\ref{fig:L100p33}, where we repeat the simulations, under the
same conditions of Fig.~\ref{fig:L50p33}, for a larger system
($L=100$).

In order to test out the robustness of these results, we have varied
the parameters controlling the protocol of the random small
perturbations. For instance, by increasing the maximum number of
different perturbations from 3 to 6 at each presentation, and/or by
changing the maximum time interval between consecutive perturbations
from $100$ to $10$, we observe the same qualitative behavior, with a
faster decrease of the correlations --- as expected, since in both cases
the noise has been increased.

\begin{figure}[!hbt]
\begin{center}
\includegraphics[width =13cm,height=7.5cm]{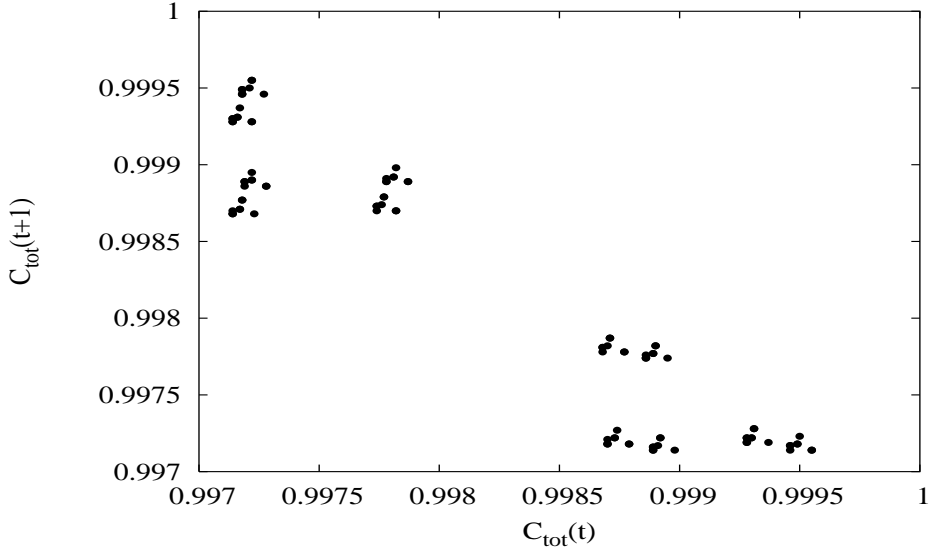}
\caption{Return map for maxima of total correlation function, for $L=50$, $t_{w}=10000$, $\Delta t_1=1000$ 
and $\Delta t_2=10000$.}
\label{spr1000-tw10000}
\end{center}
\end{figure}

What happens when the system's autocorrelations decrease due to the
small random perturbations after it has reached a periodic attractor?
Is the system driven to another cycle? To answer this question we have
studied the stability of the cycles using the following procedure: we
let the system evolve without perturbation towards its attractor for
$t_{w}=10000$ time steps, after which we perturb the system with
random small perturbations during a time interval $\Delta t_1$. Then
we turn off the perturbations and allow the system to relax during
another time interval $\Delta t_2$ after which we obtain the return
map of the correlations in the following 200 time steps.  If we
produce only one perturbation at $t_w$ ($\Delta t_1=1$) the system
remains in its original cycle, yielding a return map similar to that
of Fig.~\ref{fig:x26r}. Then, under the same conditions we repeat the
simulations adopting now $\Delta t_1=1000$ and $\Delta t_2=10000$. We
observe in Fig.~\ref{spr1000-tw10000} that the return map has changed
slightly when compared to Fig.~\ref{fig:x26r}. In particular, it no
longer shows $C_{tot}=1$ in the time series, but remains periodic,
signaling that the system has shifted to a different cycle due to the
perturbations. Apparently the periods observed in Figs~\ref{fig:x26r}
and~\ref{spr1000-tw10000} are the same. The distribution of periods
and transient times are currently under investigation, results will be
published elsewhere. Fig.~\ref{spr1000-tw10000} remains the same by
increasing $\Delta t_2$, which guarantees that the differences with
respect to Fig.~\ref{fig:x26r} are not a transient effect.  

According to these results, the role the noise plays, if allowed
enough time to perturb the system significantly, is to drive the
system from one attractor to a nearby one, which suggests that there
is a family of periodic attractors which can be very close to one
another in phase space. These effects, however, take place on a time
scale which is longer [${\cal O}(10^3-10^4)$ time steps] than the
lifetime of the mice [${\cal O}(10^2-10^3)$ time steps]. The small
perturbations are therefore of little importance to the CA dynamics
in the biologically relevant time scale, as suggested in previous
work~\cite{zsb98}.

\begin{figure}[!tb]
\begin{center}
\includegraphics[width = 0.5\textwidth, angle=-90]{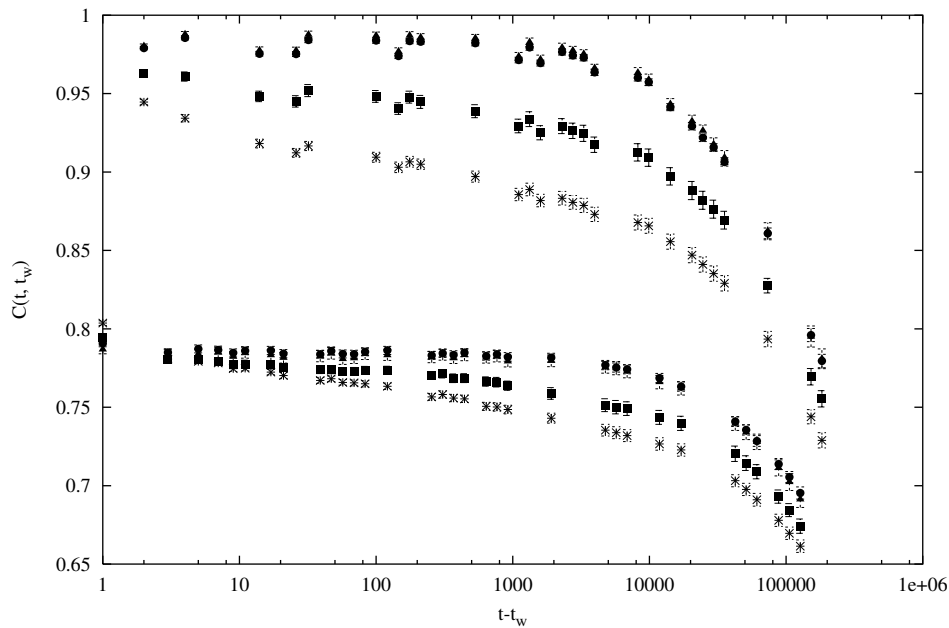}
\caption{$C_{22}(t+t_{w};t_{w})$ for different waiting times before small perturbations.
 From bottom to top, $t_{w}=100$, 5000, 10000.}
\label{fig:aging}
\end{center}
\end{figure}

The behavior of the autocorrelation functions in
Figs.~\ref{fig:L50p33} and~\ref{fig:L100p33} is reminiscent of what is
observed in glassy systems. Here the CA approaches a periodic
attractor, being thereafter deflected by the small perturbations. In
glassy systems, temperature drives the system from one local minimum
to another, preventing it from getting stuck in local minima of the
potential energy surface. As the {\em physical age} of the system
(characterized by the value of the waiting time $t_w$) grows, it finds
itself exploring deeper and deeper regions of a rough potential energy
surface, diffusing towards equilibrium~\cite{dest}.  Due to the
roughness of the potential energy, equilibrium is only attained on
very long time scales and relaxation is very slow. The older the
system is the more it gets confined to a restricted region of phase
space and the time scales for relaxation get longer and longer.
Eventually, if we wait long enough, the system equilibrates and the
dynamics becomes stationary, losing sensitivity to the waiting time.
This picture of an aging physical system is reminiscent to the loss of
plasticity for adaptation in a living organism as it gets older.  In
either case, one observes a strong dependence on the waiting time
$t_w$, evident when measuring two-time quantities like autocorrelation
functions and responses. Results for the CA model are shown in
Fig.~\ref{fig:aging}, where we see the decay of autocorrelations for
three systems with different {\em ages} or waiting times. Note that
the horizontal axis is the time difference between the total time and
the waiting time. The three curves should collapse in the case that
the dynamical evolution is stationary.

\subsection{Large immunizations}
\label{sec:large}

What is the role of the large perturbations on the dynamics of the
system?  From previous studies we would expect the large perturbations
to accelerate the aging process: while the random small perturbations
would change the route to the natural attractor of the system, the
large ones would reduce the transient time, a conjecture that would
explain the loss of plasticity of the older mice~\cite{zsb98}. The
protocol adopted is the one described in Section~\ref{sec:themodel},
with six-layer perturbations every 35 time steps, always at the same
sites.

\begin{figure}[!tb]
\begin{center}
\includegraphics[width = 0.5\textwidth, angle=-90]{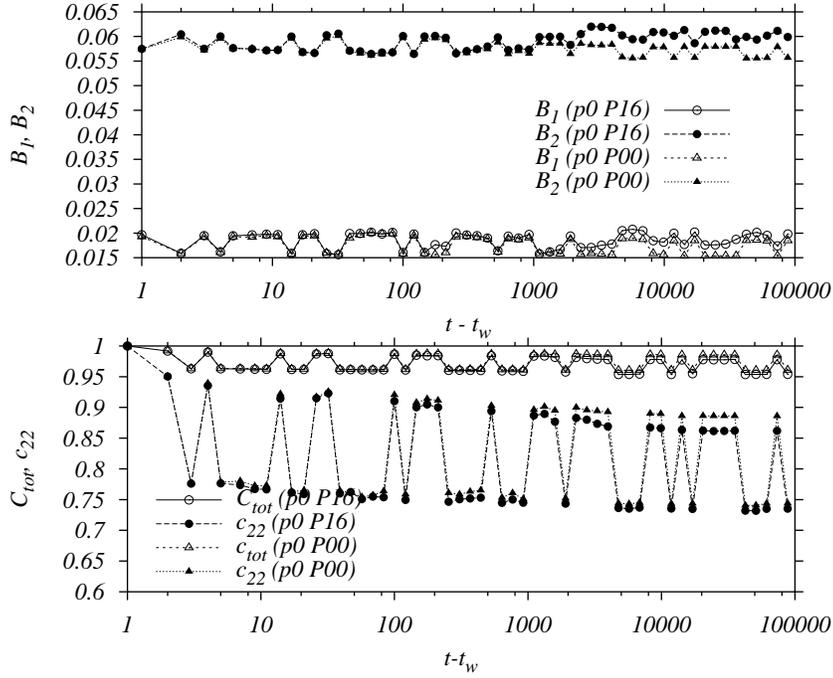}
\caption{Densities vs. time (upper panel) and autocorrelations vs. $t-
t_w$ for a system with large immunizations, $L=50$ and
$t_w=100$. Triangles: without perturbations of any kind. Circles: both
large immunizations and small perturbations.}
\label{fig:x26p0P16a0tw100}
\end{center}
\end{figure}

In Fig.~\ref{fig:x26p0P16a0tw100} we compare the results obtained for
the unperturbed system and those of the system subjected only to large
immunizations starting at $t_w$ (note that the densities are now also
plotted as functions of $t-t_w$). Somewhat surprisingly, the decrease
of the correlations for large perturbations is small, when compared to
the case of the small perturbations (compare with
Fig.~\ref{fig:L50p33}). In hindsight, however, this can be understood
because the large perturbations are always produced {\it in the same
region\/}. For very long times the correlations will attain a
stationary regime whose active populations contain at least part of
the region involved in the immunizations~\cite{zsb98}. In other words,
the driving produced by the large periodic perturbations is of a
completely different nature than that of the small perturbations. The
large perturbations seem to play a selective role: the cycles that the
system can reach are restricted to those that contain at least part of
the populations incorporated during the immunization.  According to
this picture, the aging effects observed in Ref.~\cite{zsb98} could be
simply related to the exploration of the phase space: the older the
system the less possibilities of choosing new cycles it would have. In
Section~\ref{sec:discussion} we discuss this issue further.

\begin{figure}[!tb]
\begin{center}
\includegraphics[width = 0.5\textwidth,angle=-90]{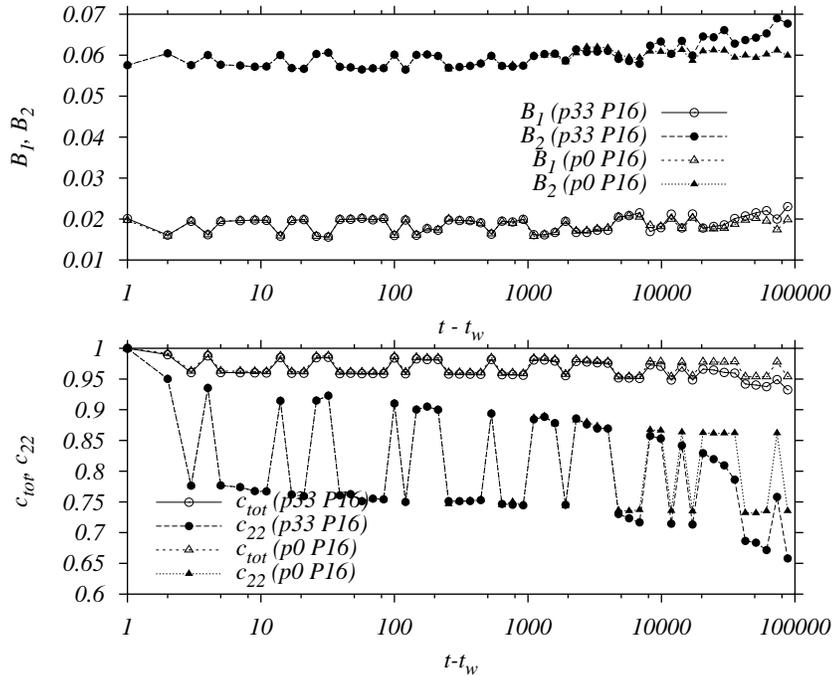}
\caption{Densities of active populations vs. time (upper panel) and autocorrelations vs. $t-
t_w$ for $t_w=100$ and $L=50$. Triangles: large immunizations
only. Circles: both large immunizations and small perturbations.}
\label{fig:x26p33P16a0tw100}
\end{center}
\end{figure}

In Fig.~\ref{fig:x26p33P16a0tw100} we compare the time evolution of
the densities $B_1$ and $B_2$ and the autocorrelation functions for
the system subjected only to large perturbations and for the system
with both small perturbations and large immunizations.  The results,
as expected, confirm the dominance of the small perturbations, over
the large ones, in driving the system faster to a different
attractor. The increase of the densities around $t\sim 10^4$ for
$L=50$ corresponds to the same finite size effects occurring in
Fig.~\ref{fig:L50p33}, being associated to the active populations
which have been incorporated into the network by means of the
immunization protocol.

\section{Concluding Remarks}
\label{sec:discussion}

We have analyzed the dynamics of a cellular automata model for the
B-cell repertoire which has reproduced experimental results of
immunizations on mice. Our analyses provide a broader context in which
memory and plasticity take place, as discussed in Ref.~\cite{zsb98}.

Defining quantities analogous to the autocorrelation functions used in
glassy systems, we have shown that the biologically relevant phenomena
takes place in the transient regime of the model. The dependence of
these functions on both $t$ and $t_w$ (as opposed to the difference
$t-t_w$ only) is reminiscent of the ``aging'' behavior observed in
glassy systems, despite the fact that the underlying dynamics of both
systems are controlled by completely different mechanisms. 

Starting from different initial conditions the deterministic dynamics
takes the system towards a family of long-period attractors.  When
subjected to random small perturbations (Fig~\ref{fig:dano}), the
system is driven towards a new attractor of the family, revealing that
most of the noise is assimilated. However, only part of the large
perturbations is incorporated, due to the mechanisms of activation and
suppression, leading to a saturation of the learning
process~\cite{zsb98}.

From the biological point of view, the history of the mouse (sample)
will be written by the different antigen presentations (random
perturbations), starting from its ``birth'' (initial condition). Since
the system is large but finite there is a maximum amount of
information it may incorporate. The closer it is to its ``destiny''
attractor, the less information it is able to learn, since the deeper
it already is in a given basin of attraction. Therefore the biological
aging corroborated by the experimental results may be simply a
consequence of this dynamical feature.

\begin{figure}[!tb]
\begin{center}
\includegraphics[width =7.5cm,height=12cm, angle=-90]{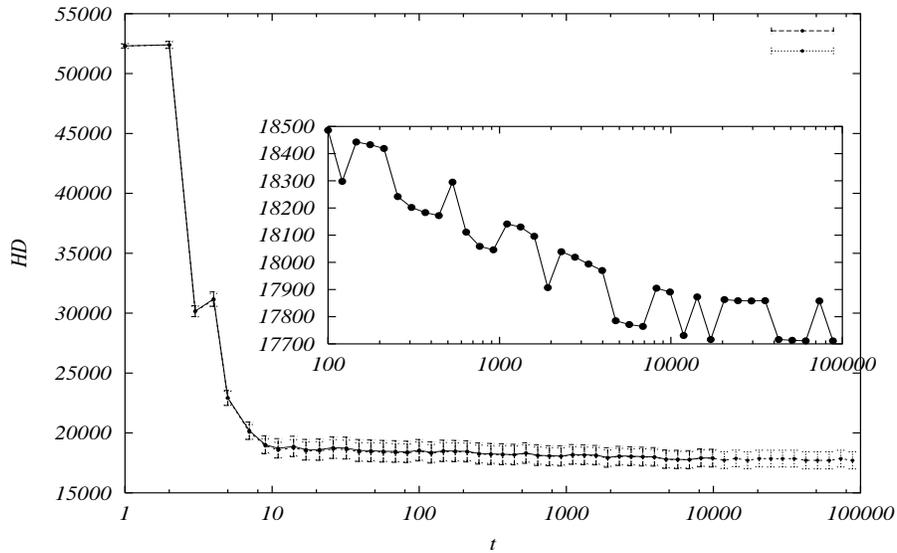}
\caption{Hamming distance between 100 pairs of random configurations as
a function of time (average and standard deviation). Inset: evolution
for $t>100$ shows that the HD attains a stationary regime for
sufficiently long times (standard deviation not shown).}
\label{fig:plotdivergence}
\end{center}
\end{figure}

From the results obtained up to now there are evidences that the
purely deterministic dynamics is therefore non-ergodic. Despite their
large number, however, the family of periodic attractors occupy only a
fraction of the phase space. Evidences of this compression in phase
space has been obtained in the following computer experiment:
selecting randomly two initial configurations (with the same initial
concentration), we measured the Hamming distance between them as a
function of time as both systems evolve without any perturbation. In
Fig.~\ref{fig:plotdivergence} we show the evolution of the average HD
for $100$ pairs (error bars are standard deviations). During the first
10 time steps the HD decreases very rapidly, but for $t>100$ we
observe a quasi-stationary regime, reflecting the slow driving to the
attractors (see inset). Note that this behavior is somewhat similar to
that of the autocorrelation functions for small $t_w$.

\begin{figure}[!tb]
\begin{center}
\includegraphics[width = 0.5\textwidth, angle=0]{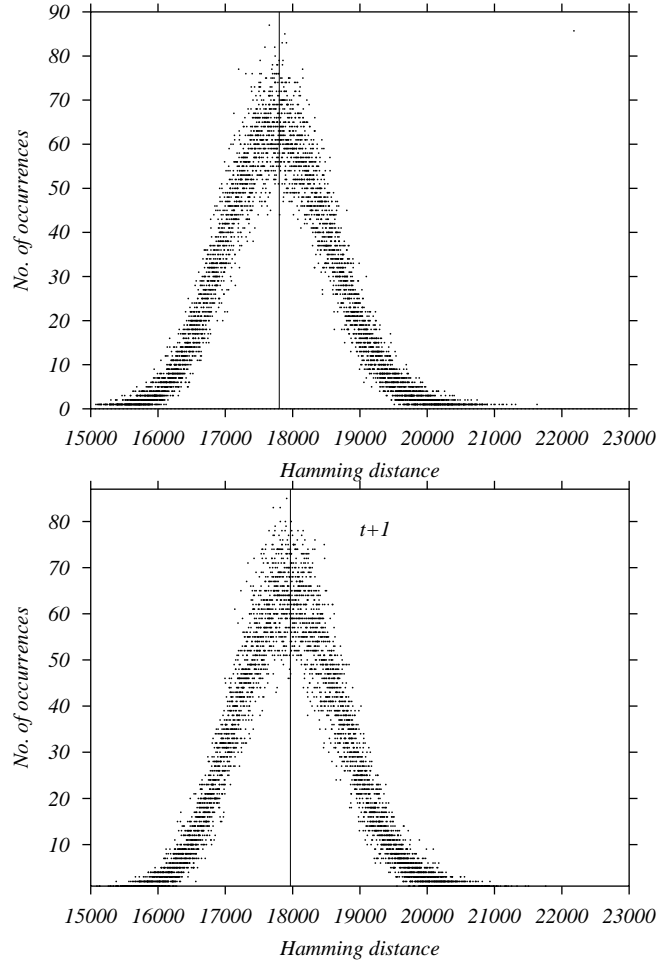}
\caption{Distribution of Hamming distance sampled over 500(500-1)/2
configuration pairs at time $t=4000$ and $t=4001$. Vertical lines show
the average value.}
\label{fig:ht4000nr500P0p0}
\end{center}
\end{figure}

Still focusing on the evidence of phase space compression, in
Fig.~\ref{fig:ht4000nr500P0p0} we present the distribution of HD
sampled from $500(500-1)/2$ pairs, for two consecutive time steps
($t=4000$ and $t=4001$). The distributions can be well described by a
Gaussian, with a width that remains approximately constant (even for
long times). Notice that the average value for $t=4001$ is slightly
larger than for $t=4000$, reflecting the oscillatory behavior of the
average HD as the pair of samples reaches their periodic
attractors. It should be noted that, for large $N$, the Central Limit
Theorem assures that randomly chosen initial configurations (with the
same $x$) naturally give rise to a Gaussian distribution of the HD
between any two of them, at time zero. Interestingly, the CA dynamics
does not change the shape of the distribution, its only effect being
to essentially shift the Gaussian towards lower mean values, on a long
time scale. The spatio-temporal structure of the cycles, as well as
their transients and basins of attraction, would be the object of
further study and will be published elsewhere.

This work was partially supported by CNPq, Faperj, Finep, Capes and
Projeto Enxoval-UFPE. MC and RMZS acknowledge IF-UFF, where part of
this work was done during their stay there.

\end{document}